\documentclass[aps,preprint,preprintnumbers,amsmath,amssymb,amsfonts,groupedaddress,superscriptaddress,showpacs,showkeys,floatfix]{revtex4-1}

\usepackage{graphicx}
\usepackage{subfigure}
\usepackage{bm}
\usepackage{textcomp}
\usepackage{color}
\usepackage{array}
\usepackage{epstopdf}

\begin{document}

\title{The study of intrinsic defect state of FeSe \\ with scanning tunneling microscopy}

\author{Kunliang Bu}
\affiliation{Zhejiang Province Key Laboratory of Quantum Technology and Device, Department of Physics, Zhejiang University, Hangzhou 310027, China}
\author{Bo Wang}
\affiliation{State Key Laboratory of Silicon Materials, School of Materials Science and Engineering, Zhejiang University, Hangzhou 310027, China}
\author{Wenhao Zhang}
\affiliation{Zhejiang Province Key Laboratory of Quantum Technology and Device, Department of Physics, Zhejiang University, Hangzhou 310027, China}
\author{Ying Fei}
\affiliation{Zhejiang Province Key Laboratory of Quantum Technology and Device, Department of Physics, Zhejiang University, Hangzhou 310027, China}
\author{Yuan Zheng}
\affiliation{Zhejiang Province Key Laboratory of Quantum Technology and Device, Department of Physics, Zhejiang University, Hangzhou 310027, China}
\author{Fangzhou Ai}
\affiliation{Zhejiang Province Key Laboratory of Quantum Technology and Device, Department of Physics, Zhejiang University, Hangzhou 310027, China}
\author{Zongxiu Wu}
\affiliation{Zhejiang Province Key Laboratory of Quantum Technology and Device, Department of Physics, Zhejiang University, Hangzhou 310027, China}
\author{Qisi Wang}
\affiliation{State Key Laboratory of Surface Physics and Department of Physics, Fudan University, Shanghai 200433, China}
\author{Hongliang Wo}
\affiliation{State Key Laboratory of Surface Physics and Department of Physics, Fudan University, Shanghai 200433, China}
\author{Jun Zhao}
\affiliation{State Key Laboratory of Surface Physics and Department of Physics, Fudan University, Shanghai 200433, China}
\affiliation{Collaborative Innovation Center of Advanced Microstructures, Nanjing University, Nanjing 210093, China}
\author{Chuanhong Jin}
\affiliation{State Key Laboratory of Silicon Materials, School of Materials Science and Engineering, Zhejiang University, Hangzhou 310027, China}
\author{Yi Yin}
\email{yiyin@zju.edu.cn}
\affiliation{Zhejiang Province Key Laboratory of Quantum Technology and Device, Department of Physics, Zhejiang University, Hangzhou 310027, China}
\affiliation{Collaborative Innovation Center of Advanced Microstructures, Nanjing University, Nanjing 210093, China}

\begin{abstract}
We apply high resolution scanning tunneling microscopy to study intrinsic defect states of bulk FeSe.
Four types of intrinsic defects including the type I dumbbell, type II dumbbell, top-layer Se vacancy
and inner-layer Se-site defect are extensively analyzed by scanning tunneling spectroscopy.
From characterized depression and enhancement of density of states measured in a large energy range,
the type I dumbbell and type II dumbbell are determined to be the Fe vacancy and Se$_\mathrm{Fe}$ defect,
respectively. The top-layer Se vacancy and possible inner-layer Se-site vacancy are also determined by
spectroscopy analysis. The determination of defects are compared and largely confirmed in the annular dark-field
scanning transmission electron microscopy measurement of the exfoliated FeSe. The detailed mapping
of defect states in our experiment lays the foundation for a comparison with
complex theoretical calculations in the future.
\end{abstract}

\maketitle

\section{INTRODUCTION}

Atomic defects are ubiquitous in condensed matter materials. The type, density and distribution of
atomic defects can be controlled in material preparation to introduce doped carriers~\cite{LeeRMP06, KamiharaJACS08},
tune phase transitions~\cite{MatsuuraNATCOMMUN17, HafiezPRB16}, pin vortices in superconductors~\cite{PanPRL00, LiuPRX18},
and provide other related applications.
The microscopic effect of atomic defects has been well studied by
atomic-resolved scanning tunneling microscopy (STM).
The defect scattering of electronic states leads to a quasiparticle interference (QPI) pattern
from which the electronic band structure of materials can be extracted~\cite{HuangPRL15, BuPRB18}.
Magnetic or nonmagnetic defects can induce a resonant in-gap state for probing pairing
symmetry of high-$T_\mathrm{c}$ superconductors~\cite{BalatskyRMP06, FischerRMP07, ZhangPRL09, LiuPRL19, ChiPRB16}.
Generally, the defect-induced change of the local density of states (DOS) includes information of the interaction
between defects and the bulk material, rendering insights about the determination of defects and material properties.

FeSe is the structurally simplest iron-based superconductor~\cite{HsuPNAS08}. The critical
temperature of the parent bulk FeSe is $T_\mathrm{c}$$\sim$ 9 K, which can be astonishingly
enhanced to much higher values by the intercalation~\cite{GuoPRB10, YingSREP12, LucasNMAT13, LuNMAT15} or the doping with K
adatoms~\cite{WenNATCOMMUN16, MiyataNMAT15, YeARXIV15, SongPRL16}.
A related high-$T_\mathrm{c}$ system is the monolayer FeSe grown
on SrTiO$_3$ substrates~\cite{WangCPL12, HeNMAT13}.
A dumbbell defect has been observed in FeSe and FeSe-related
systems, with defects as scattering centers of a QPI pattern at ultra-low
temperatures~\cite{KasaharaPNAS14, SprauSCIENCE17, KostinNMAT18, FanNPHYS15, HuangPRB16},
pinning sites of nematic order~\cite{SongSCIENCE11, SongPRL12, WatashigePRX15}
and charge order~\cite{LiNPHYS17}, and the touchstone of the paring symmetry~\cite{JiaoPRB17, DuNPHYS17, YanPRB16}.

In most previous STM reports of FeSe~\cite{SprauSCIENCE17, WatashigePRX15, JiaoSREP17, JiaoPRB17},
the voltage range of scanning tunneling spectroscopy (STS) is limited within $\pm20$ mV,
close to the superconducting gap ($\Delta\approx2.5$ mV).
Correspondingly such delicate spectroscopy can be used to clarify the pairing
symmetry of the superconductor~\cite{SprauSCIENCE17, WatashigePRX15, JiaoSREP17, JiaoPRB17}.
Despite of the insightful information reported from the intrinsic defects
such as dumbbell defect and Se vacancy~\cite{JiaoPRB17, KasaharaPNAS14, DuNPHYS17, YanPRB16, LiuNANOLETT19},
spectroscopy at the defect site varies for different experiments~\cite{JiaoPRB17, KasaharaPNAS14}.
Defect state has been extensively studied on monolayer FeSe/SrTiO$_3$ films~\cite{FanNPHYS15, LiuPRB18, LiuNANOLETT19}.
In this paper, we explore four typical types of intrinsic atomic defects in the bulk FeSe,
including the type I dumbbell, type II dumbbell, top-layer Se vacancy and inner-layer Se-site defect.
We are more interested in the STS with a large voltage range and the general defect-induced
change of DOS. From the STS results, two types of dumbbell defects are
determined to be the Fe vacancy and antisite Se$_\mathrm{Fe}$ defect.
The inner-layer Se-site defect is attributed to possible inner-layer Se vacancy.
As a complementary technique, annular dark-field scanning transmission electron microscopy (ADF-STEM) is also
applied to the exfoliated FeSe to check intrinsic defects.
Both Fe vacancy and antisite Se$_\mathrm{Fe}$ defect are detected and determined,
consistent with the STM results. We provide a detailed mapping of defect states, which lays
a foundation for comparison with complex theoretical calculations in the future.

\section{EXPERIMENTAL METHOD}
High quality single crystals of FeSe in stoichiometry were grown using the KCl-AlCl$_3$ flux technique~\cite{WangNATCOMMUN16}.
Typical size of our FeSe samples is 2 mm $\times$ 2 mm.
The crystallization was confirmed by the X-ray diffraction (XRD) measurement.
The resistance curve upon warming indicates a superconducting transition temperature of $T_c$ = 8.9 K.

The STM and STS experiments were conducted in a commercial system with ultra-high vacuum and low temperature.
The samples were cleaved at liquid nitrogen temperature at $\sim$ 77 K and were inserted into the STM head immediately.
An electrochemically etched tungsten tip was treated by e-beam sputtering and field emission on Au (111) surface before the STM measurement.
A constant current mode with a feedback loop control was used to take STM images.
The $dI/dV$ spectra were taken at a bias modulation of 2 mV with a modulation frequency of 1213.7 Hz.
Since we are mostly interested in determining the defect states, we are not bound to temperatures below the superconducting transition.
All STM and STS data were acquired at liquid nitrogen temperature ($\sim$ 77 K).
The temperature is a bit lower than the structural phase transition ($T_s$$\sim$ 90 K).

The ADF-STEM measurement was performed in a so-called probe-corrected STEM (FEI Titan Chemi STEM) operated
at 200 kV at room temperature. The convergence angle was set at 21.4 mrad and the range of acceptance
angle of ADF detector was between 53 mrad and 200 mrad.
Few layers of FeSe on silicon nitride (SiN$_x$) grid was prepared by a mechanical exfoliation.
The cleavage plane was exposed to atmosphere for less than 20 min and was not contaminated
by any other reagents.

\section{RESULTS AND DISCUSSION}

\begin{figure}[tp]
\centering
\includegraphics[width=.5\columnwidth]{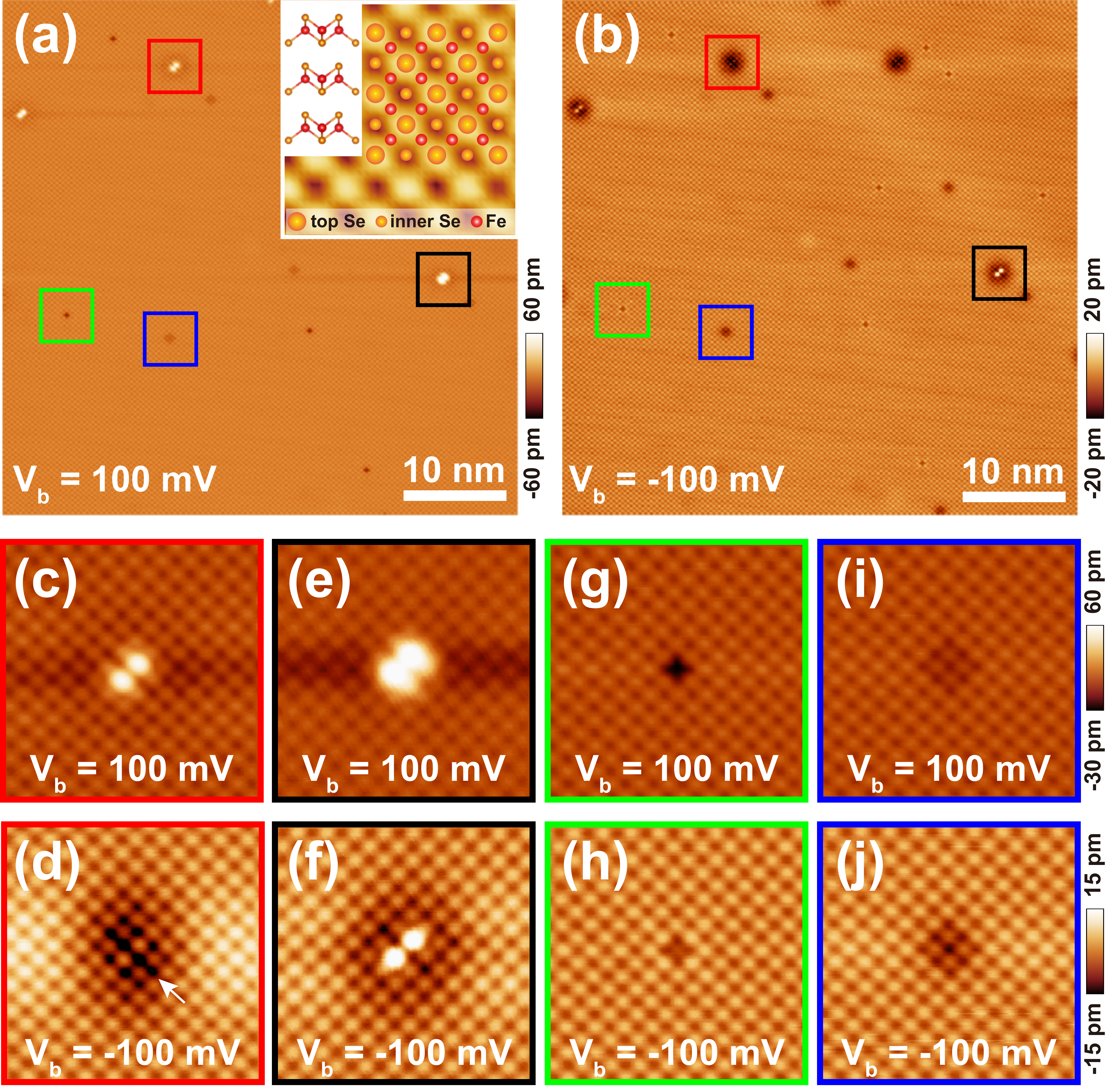}
\caption{(a, b) Atomic resolved large area topographies (50 nm $\times$ 50 nm) under positive
and negative bias voltages, respectively.
The red, black, green and blue frames label the type I dumbbell, type II dumbbell,
top-layer Se vacancy and inner-layer Se-site defect, respectively.
The inset in (a) is an enlarged image with a schematic top-view lattice superimposed on it.
The left part of the inset is the schematic front-view of FeSe structure.
The tunneling condition is $V_\mathrm{b}$ = 100 mV, $I_\mathrm{s}$ = 20 pA for (a) and
$V_\mathrm{b} = -100$ mV, $I_\mathrm{s} = 20$ pA for (b).
(c, e, g, i) Zoom-in images of the type I dumbbell, type II dumbbell, top layer Se
vacancy and inner-layer Se-site defect, respectively. The tunneling condition is $V_\mathrm{b} = 100$ mV,
$I_\mathrm{s} = 20$ pA.
(d, f, h, j) Zoom-in images of the same area in (c, e, g, i). The tunneling condition is
$V_\mathrm{b} = -100$ mV, $I_\mathrm{s} = 20$ pA.
}
\label{fig_n01}
\end{figure}

The unit structure of FeSe
is composed of a square iron plane sandwiched by two square Se planes~\cite{HsuPNAS08}.
The bulk single crystal of FeSe can be cleaved between two adjacent Se layers,
with an electrically neutral Se layer exposed for the STM measurement.
Figures~\ref{fig_n01}(a) and~\ref{fig_n01}(b) display two atomic-resolved topographies of the same area
under positive and negative bias voltages, respectively.
Each bright spot in the topography represents a Se atom, which form a square
net of Se lattice with a lattice constant of $a_0\approx0.37$ nm.
The Fe plane and inner-layer Se plane are not discernable in topographies.
A small defect-free area topography is enlarged and shown in the inset of Fig.~\ref{fig_n01}(a),
on top of which a schematic top-view lattice is superimposed. Each iron atom is shown to be at the
bridge site between two neighboring top-layer Se atoms. Each inner-layer Se atom is at
the hollow site of the top-layer Se lattice.

Within the top-layer Se lattice, different types of defects can be observed.
We mainly focus on four types of typical defects, the type I dumbbell, the type II dumbbell,
the top-layer Se vacancy and the inner-layer Se-site defect. With the sample being an asgrown compound
of bulk FeSe, these typical defects are attributed to intrinsic defects of FeSe.
As shown in Figs.~\ref{fig_n01}(c-f), both types of dumbbell defects are centered at the Fe site.
For the type I dumbbell, the topography under positive bias voltage shows two
bright lobes on adjacent top-layer Se sites [Fig.~\ref{fig_n01}(c)].
The topography under negative bias voltage shows much suppressed lobe structure,
and a dark feature perpendicular to the lobe direction [Fig.~\ref{fig_n01}(d), white arrow].
For the type II dumbbell, the lobe structure under positive bias voltage is
brighter than that of the type I dumbbell [Fig.~\ref{fig_n01}(e)].
The topography under negative bias voltage
shows a suppressed but still bright lobe structure [Fig.~\ref{fig_n01}(f)].
The dumbbell defects are grouped into the type I and type II dumbbells, depending on
their distinct topographies under negative bias voltages.
These two types of dumbbell defects can also be distinguished from the published literatures about FeSe~\cite{JiaoSREP17, LiuPRB18}.
Besides the dumbbell defects which are located at the Fe site,
we also find defects which are located at the Se site.
With a missing Se atom in the Se lattice in Figs.~\ref{fig_n01}(g-h), the third type of defect is
determined to be the top-layer Se vacancy.
Figures~\ref{fig_n01}(i-j) show
that the fourth type of defect is centered at the inner-layer Se site while no missing atoms are found in the
top-layer Se lattice. For the present, it is assigned as an inner-layer Se-site defect.
Different from the C$_2$ symmetry shown in the topographies of dumbbell defects,
a C$_4$ symmetry is observed around both Se-site defects.
In addition, topography of both Se-site defects under positive bias voltage
is similar to that under negative bias voltage.

\begin{figure}[tp]
\includegraphics[width=.6\columnwidth]{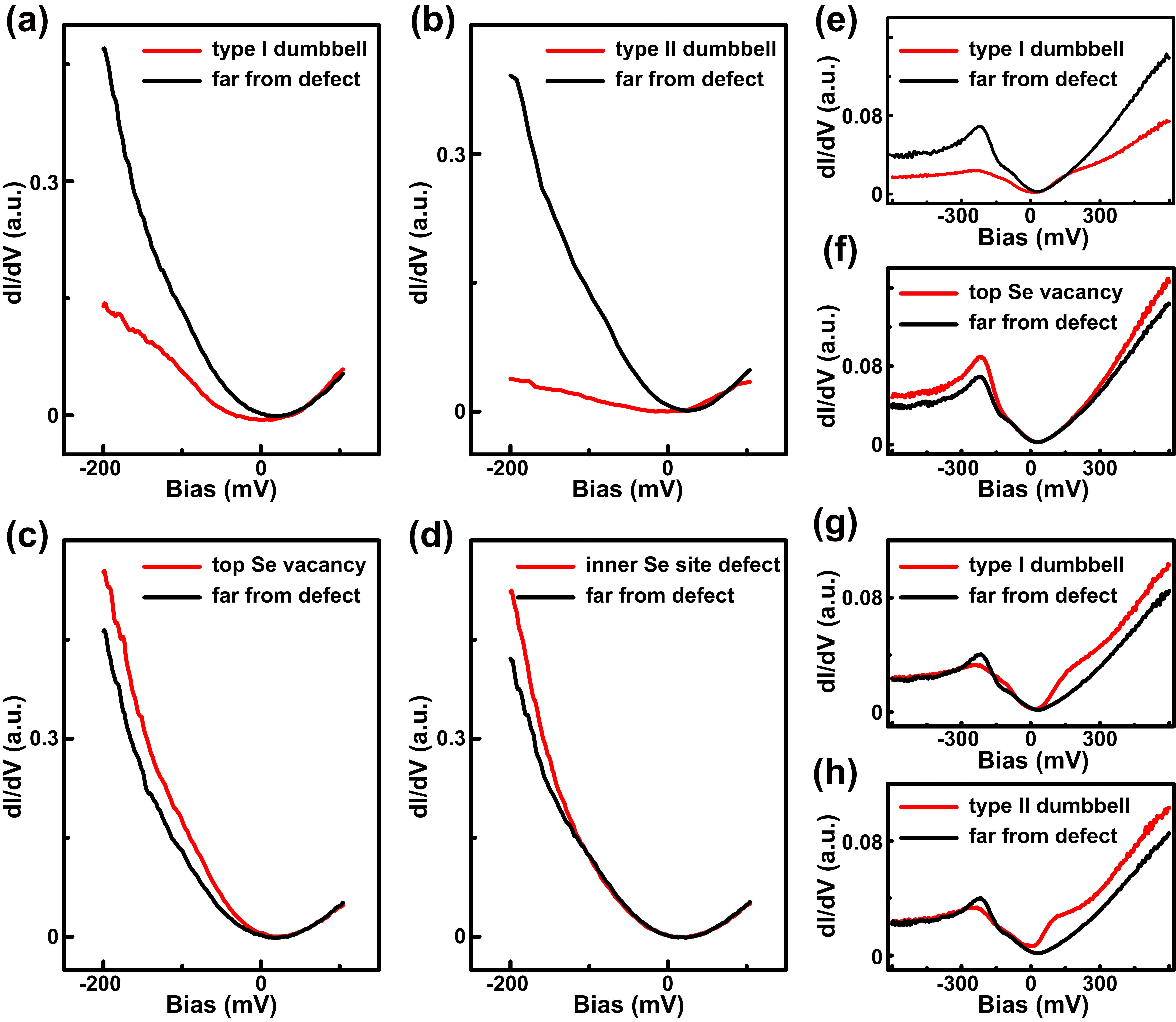}
\caption{
The $dI/dV$ spectra at the defect site (red curves) and far away from the defect site (black curves).
(a-d) are spectra for the type I dumbbell, type II dumbbell, top-layer Se vacancy and
inner-layer Se-site defect, respectively. The tunneling condition is $V_\mathrm{b} = 100$ mV and $I_\mathrm{s} = 100$ pA.
(e, f) The $dI/dV$ spectra of the type I dumbbell and top-layer Se vacancy with a
large voltage range. The tunneling condition is $V_\mathrm{b} = 100$ mV and $I_\mathrm{s} = 20$ pA.
(g, h) The $dI/dV$ spectra of the type I dumbbell and the type II dumbbell with a
large voltage range. The tunneling condition is $V_\mathrm{b} = -600$ mV and $I_\mathrm{s} = 600$ pA.
}
\label{fig_n02}
\end{figure}

The $dI/dV$ spectrum is next measured at different sites, which is proportional to the local electronic DOS.
Initially we choose a tunneling condition of $V_\mathrm{b}=100$ mV and $I_\mathrm{s}=100$ pA.
At clean area without any defects, the $dI/dV$ spectrum displays a V-shaped form
around the Fermi level, as shown by the black curve in Fig.~\ref{fig_n02}a.
In the V-shaped spectrum, the DOS at negative energy is larger than the DOS at positive energy, representing a partial particle-hole asymmetry.
At the center of the type I dumbbell defect, the $dI/dV$ spectrum is measured and shown
by the red curve in Fig.~\ref{fig_n02}(a). With the tunneling bias
voltage set at $V_\mathrm{b}=100$ mV, the DOS at positive energy is similar to that
of the clean-area spectrum. The DOS at negative energy is relatively depressed,
displaying a more symmetric DOS around the Fermi level.
For the type II dumbbell defect, the DOS at negative energy is further depressed to
be smaller than the DOS at positive energy, as shown by the red curve in Fig.~\ref{fig_n02}(b).
Figure~\ref{fig_n02}(c) shows the $dI/dV$ spectrum of the top-layer Se vacancy.
Compared with the clean-area spectrum, the DOS at negative energy is enhanced, instead of
being depressed like that in dumbbell defects. For the inner-layer
Se-site defect, the $dI/dV$ spectrum shows a similar but smaller enhancement
of particle-hole asymmetry [Fig.~\ref{fig_n02}(d)]. The inner-layer Se-site defect is
most possibly an inner-layer Se vacancy, with the subtle spectral difference
attributed to a larger distance between the defect and the tip.

The asgrown bulk FeSe is a bad metal. Not like for a semiconductor, the defect-induced
change of DOS cannot be explained by a simple argument, or calculated by general density
functional theory because of the strong-correlation effect. For a stoichiometric FeSe
single crystal, the iron site defect could probably be the Fe vacancy (V$_\mathrm{Fe}$) or
an antisite defect of Se$_\mathrm{Fe}$, with the symbol Se$_\mathrm{Fe}$ denoting a
Se atom occupying a site that should have had a host Fe atom on it. The Fe vacancy can
be recognized as a substitution of a high valence state ion (Fe$^{2+}$) with a low
valence state ion (V$_\mathrm{Fe}^{0+}$). For the antisite defect Se$_\mathrm{Fe}$,
Fe$^{2+}$ is substituted by an even lower valence state ion (Se$^{2-}$). For the
Se vacancy, a low valence state ion (Se$^{2-}$) is substituted by a high valence
state ion (V$_\mathrm{Se}^{0+}$). With the different valence change of defects,
the scattering potential with different degrees may depress and enhance the DOS
at negative energy correspondingly. Based on the order of valence change, the
type I and type II dumbbell defects are determined to be the Fe vacancy and
Se$_\mathrm{Fe}$ defect, respectively. In addition, the approximate density
of the type I dumbbell defect is observed to decrease with the experimental progress or the cooling
cycles, consistent with the fact that the Fe vacancy can be wiped out by annealing~\cite{HuangNANOLETT16}.

The defect-induced change of DOS can be explored in a larger voltage range around the Fermi
level. For the type I dumbbell defect, the $dI/dV$ spectrum from -600 mV to 600 mV is measured
and compared with a clean-area spectrum [Fig.~\ref{fig_n02}(e)]. With a bias voltage
$V_\mathrm{b}=100$ mV, the tunneling current is changed to $I_\mathrm{s}=20$ pA to avoid
an overload of the measured lock-in signal. Consistent with that in Fig.~\ref{fig_n02}(a), the DOS
at negative energy is strongly depressed, including the peak around -220 mV in the clean-area
spectrum. On the other hand, the DOS at positive energy, from 200 mV to 600 mV, is also depressed.
For the top-layer Se vacancy, the similar large-range spectrum is measured and shown in
Fig.~\ref{fig_n02}(f). Consistent with that in Fig.~\ref{fig_n02}(c), the DOS at negative
energy is overall enhanced. For two types of dumbbell defects, the tunneling condition is
also changed to $V_\mathrm{b}=-600$ mV and $I_\mathrm{s}=600$ pA for a different measurement
of the $dI/dV$ spectrum. As shown in Figs.~\ref{fig_n02}(g) and \ref{fig_n02}(h), a protruding kink
below 200 mV is the main dumbbell-induced change of DOS at positive energy (compared with the clean-area $dI/dV$ spectrum), which is
more obvious in Fig.~\ref{fig_n02}(h) for the type II dumbbell defect. The relative particle-hole
asymmetry around the Fermi level is maintained in the large-energy-range $dI/dV$ spectra.
Defect-induced $dI/dV$ spectra have been reproducibly tested at different
samples and with different tips.

\begin{figure}[tp]
\includegraphics[width=.9\columnwidth]{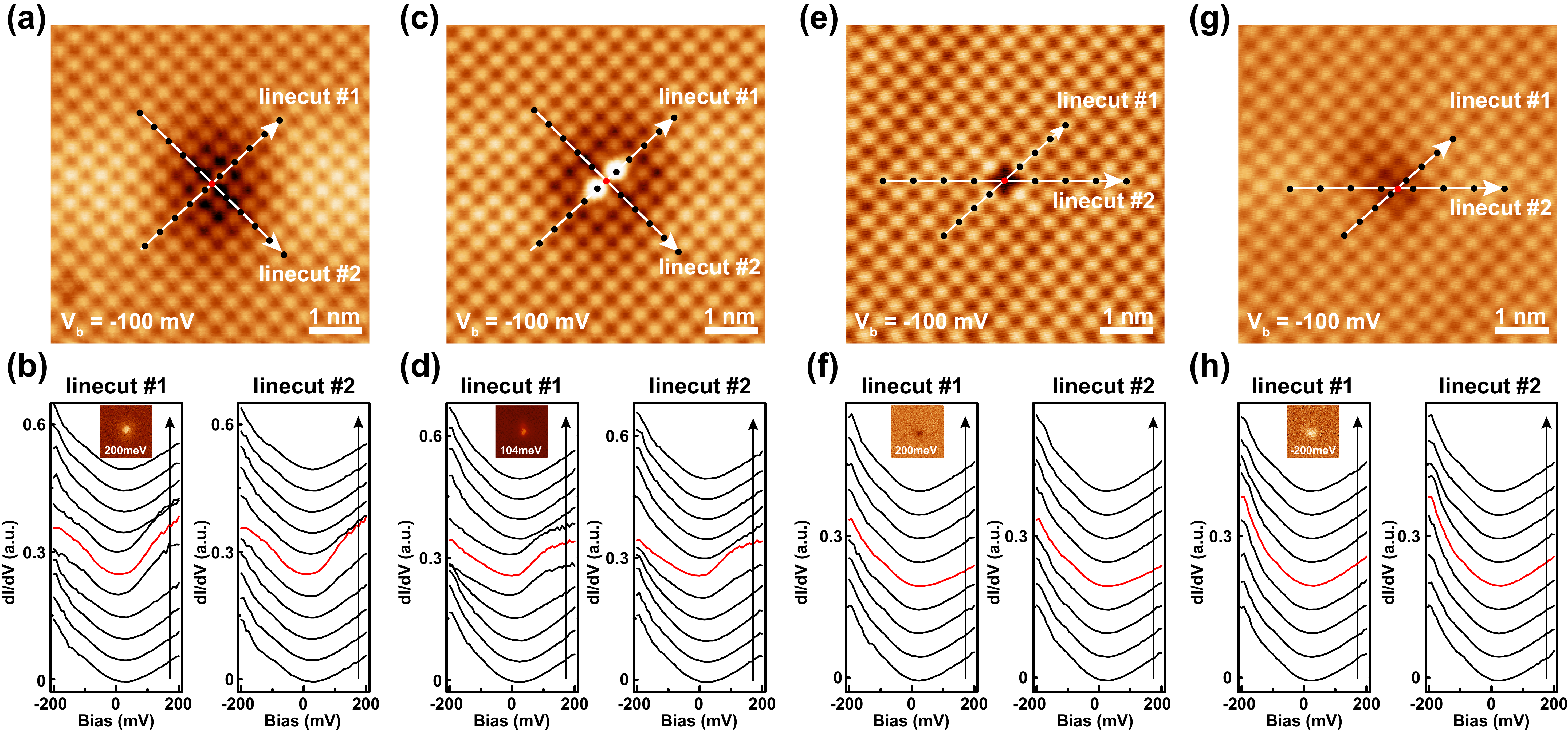}
\caption{
(a) Topography of the type I dumbbell. Two white arrows indicate linecut $\#1$ and linecut $\#2$, respectively. Black and red dots are positions where the spectra were taken.
(b) Left and right panels show series of spectra along the linecut $\#1$ and linecut $\#2$, respectively.
The inset of (b) is the conductance map of 200 mV around the type I dumbbell.
(c-h) The same as (a) and (b) but with topographies and spectra for the type II dumbbell (c, d), the top-layer
Se vacancy (e, f) and the inner-layer Se-site defect (g, h). All the topographies and spectra
were taken under $V_\mathrm{b} = -100$ mV and $I_\mathrm{s} = 100$ pA. The insets of (d), (f) and (h)
are conductance maps of 104 meV, 200 meV and -200 meV around the type II dumbbell, top-layer Se vacancy and inner-layer Se-site defect, respectively.
The conductance maps in (f) and (h) are selected with different signs to present the defect states more clearly.
}
\label{fig_n03}
\end{figure}

We then explore the spatial distribution of the electronic states around different types of defects.
Series of linecuts of $dI/dV$ spectra are measured across defects. The tunneling
condition is chosen to be $V_\mathrm{b}=-100$ mV and $I_\mathrm{s}=100$ pA.
Due to the negative bias voltage applied to this measurement, the spectra are normalized by the DOS of occupied states.
In Figs.~\ref{fig_n03}(b,d,f,h), the red curves are the $dI/dV$ spectra taken at the center of each type of defect,
consistent with the spectra in Fig.~\ref{fig_n02}. For dumbbell defects,
linecut $\#1$ is along the dumbbell direction and linecut $\#2$ is perpendicular to the dumbbell direction,
as labeled on the topographies in Figs.~\ref{fig_n03}(a) and~\ref{fig_n03}(c).
At two Se sites adjacent to the defect site in linecut $\#1$, the $dI/dV$ spectra are still
similar to that at the defect center, despite the bright lobe structure at these sites in topographies.
The change of DOS fades away at next neighboring Se-sites,
with a distance of 1.5 Se-Se lattice ($\sim$ 0.57 nm) from the defect site. Similar to that along
linecut $\#1$, the change of DOS along linecut $\#2$ fades away with a similar distance away from the defect site.
The spatial distribution of the defect state is illustrated in the conductance map of 200 mV,
as shown in the inset of Fig.~\ref{fig_n03}(b).
The type II dumbbell shares a similar enhancement of DOS at positive voltages as the type I dumbbell,
as shown in Figs.~\ref{fig_n03}(c) and~\ref{fig_n03}(d). The DOS at the protruding kink
below 200 mV is selected as a defect-induced signal, and the conductance map of 104 meV
is shown in the inset of Fig.~\ref{fig_n03}(d). For the monolayer FeSe interfaced with
SrTiO$_3$, the DFT calculation reveals that the two protruding Se orbitals around the Fe
vacancy lead to the bright lobes of a dumbbell~\cite{HuangNANOLETT16}. For the Fe
vacancy and antisite Se$_\mathrm{Fe}$, our detailed mapping of the defect state
suggests that the neighboring Se orbitals should lead to the similar but different
lobes of dumbbell. The distribution of the defect state is however relatively isotropic
around the center, and quickly fades away with increasing distance from the defect center.
We note that there are unidirectional dark stripes straddle each dumbbell defect at temperatures much lower than $T_s$~\cite{SongPRL12, WatashigePRX15, JiaoSREP17, SprauSCIENCE17}.
Because the measurement temperature in our experiment is at $77$ K, only a bit lower than $T_s$,
the more long-ranged unidirectional depressions are less discernable in our topographies.

For the C$_4$ symmetric top-layer Se vacancy and the inner-layer Se vacancy,
the results are similarly measured and displayed.
As shown in the topographies in Figs.~\ref{fig_n03}(e) and~\ref{fig_n03}(g),
linecut $\#1$ is along the Se-Se lattice direction and linecut $\#2$ is along the $45^{\circ}$ direction
with respect to the lattice direction. For the top-layer Se vacancy, the
depression of DOS at 200 mV is selected as the defect-induced signal. For the inner-layer
Se-site defect, the enhancement of DOS at -200 mV is selected as the defect-induced signal.
The corresponding conductance maps are presented in insets of Figs.~\ref{fig_n03}(f) and
\ref{fig_n03}(h), with the defect state also quickly fading away from the defect
center.

\begin{figure}[tp]
\includegraphics[width=.5\columnwidth]{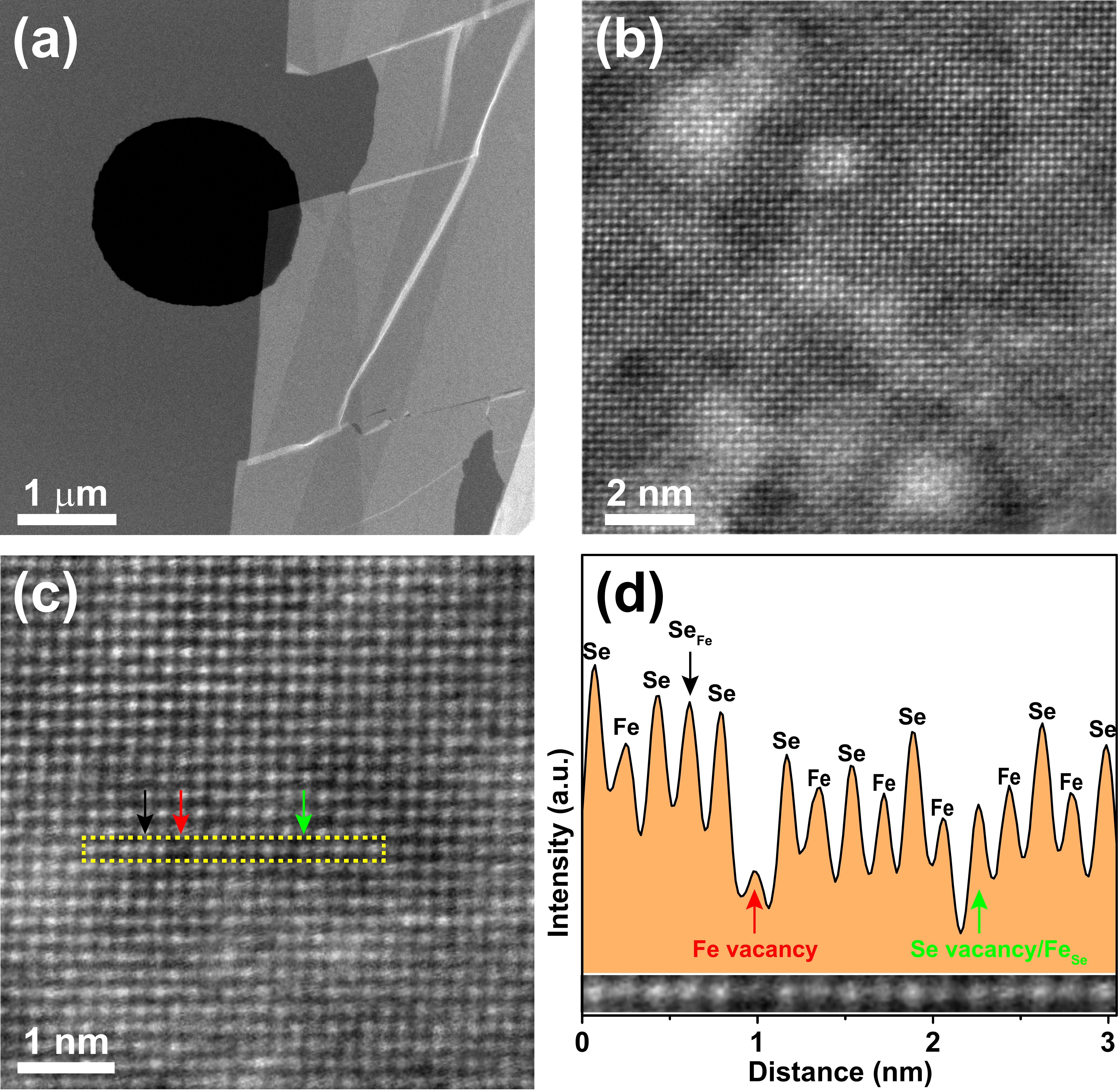}
\caption{
(a) An over view of the exfoliated FeSe on silicon nitride grid.
A three-layer FeSe is selected for the ADF-STEM measurement.
(b) A 11.9 nm $\times$ 11.9 nm ADF-STEM image of FeSe.
(c) A 5.5 nm $\times$ 5.5 nm zoom-in image. The Se$_\mathrm{Fe}$ defect, Fe vacancy
and possible Se vacancy are labeled by the black, red and green arrows, respectively.
(d) An intensity plot along atoms in the yellow dashed frame in (c). The corresponding atoms and defects are labeled.
The bottom of (d) is the zoom-in image of the yellow dashed frame in (c).}
\label{fig_n04}
\end{figure}

An ADF-STEM measurement of FeSe is further carried out for a comparison with STM results.
Figure~\ref{fig_n04}(a) shows an overview of an exfoliated FeSe on a silicon nitride (SiN$_x$) grid.
The thickness of the exfoliated FeSe could be estimated from the contrast between the SiN$_x$ substrate
and the few-layer FeSe. Because the terminal FeSe layer is easy to be oxidized when exposed to
atmosphere, we intentionally chose a three-layer FeSe,
in which the intermediate FeSe can be protected by terminal FeSe layers. In the ADF-STEM measurement,
electrons pass through the
sufficiently thin specimen of three-layer FeSe within the round hole for the collection of signal.
Figure~\ref{fig_n04}(b) shows an atomic resolved ADF-STEM image, which is an integrated
signal from the area with three layers of FeSe. The white patches in the image correspond to
oxidized amorphous FeSe on the surface, while the lattice of intermediate FeSe can still be resolved.

In a relative homogeneous area, the integrated ADF-STEM image includes both the Se and Fe lattices.
The contrast of atoms in ADF-STEM image is related with the atomic number $Z$. Then the brighter
and less bright atoms are determined to be Se and Fe atoms, respectively.
Figure.~\ref{fig_n04}(c) shows a zoom-in image with periodic bright contrast.
In a selected row of atoms [yellow dashed frame in Fig.~\ref{fig_n04}(c)],
the overall periodic bright contrast is compatible with the Se-Fe-Se atomic chain.
Here we cannot distinguish the top-layer Se and inner-layer Se atoms from the ADF-STEM image.
There are also several anomalies on the contrast of atoms, with positions labeled by the colored arrows.
The abnormal atomic contrast can be seen more clearly in the intensity profile along the atomic row,
as shown in Fig.~\ref{fig_n04}(d).
The overall higher intensity on the left portion of the profile in Fig.~\ref{fig_n04}(d) is due
to the oxidized amorphous FeSe layer, which has no influence on our analysis.
The black arrow labels a Fe site where there should be a rather low intensity
but with a similar intensity to the Se site.
This anomaly is judged to be caused by a Se$_\mathrm{Fe}$ defect.
The red arrow labels a Fe site with much lower intensity than the normal Fe site.
This anomaly is most possibly caused by the Fe vacancy.
The observation of both the Se$_\mathrm{Fe}$ defect and Fe vacancy in
ADF-STEM images is compatible with the STM results.
The green arrow labels a Se site where there should be a rather high intensity
but with a similar intensity to the Fe site.
This defect may be a Se vacancy or a substitution of Se with Fe (Fe$_\mathrm{Se}$ defect).
With the oxidized layer hindering a quantitative analysis of the intensity contrast,
we cannot distinguish the certain type of this defect only from ADF-STEM image.
Since we only find Se vacancies in STM experiment,
we prefer it to be a Se vacancy based on our STS analysis.
Stronger evidence should be proposed in the future to clarify its certain type.

\section{Summary}
We have systematically analyzed the electronic state of intrinsic defects in bulk FeSe.
For two types of Fe-site defects, the distinct topographies under negative
bias voltage lead to the type I and type II dumbbell defects. Together with
two Se-site defects, we carefully probe their spectra within a large energy range.
The different valence change of defects is related with characterized depression and
enhancement of DOS around the Fermi level, based on which the
type I and type II dumbbell defects are determined to be the Fe vacancy and
Se$_\mathrm{Fe}$ defect, respectively. This determination of defects is
compared and largely confirmed with the ADF-STEM technique.
The distributions of the defect states are further explored, showing a relatively
isotropic distribution and a quick fading away from the defect center. The detailed mapping
of the defect states can be compared with complex theoretical calculations
in the future. The determination of two different types of dumbbell defects will also
help to clarify the in-gap state~\cite{JiaoPRB17, KasaharaPNAS14} and
paring symmetry problem at an ultra-low temperature,
which is beyond our current technical capability and the scope of this work.

\begin{acknowledgments}

This work was supported by the National Basic Research Program of China (2015CB921004),
the National Natural Science Foundation of China (NSFC-11374260),
and the Fundamental Research Funds for the Central Universities in China.
The work at Fudan University was supported by the Innovation Program of Shanghai Municipal Education Commission
(2017-01-07-00-07-E00018), the National Natural Science Foundation of China (NSFC-11874119),
the National Basic Research Program of China (2015CB921302), and the National Key R\&D Program of the
MOST of China (2016YFA0300203).
This work made use of the resources of the Center of Electron Microscopy of Zhejiang University.

\end{acknowledgments}

\end{document}